\pdfoutput=1
\documentclass[preprint, review,12pt]{elsarticle}
\usepackage{float}
\usepackage{tikz}
\usetikzlibrary{arrows.meta}
\usepackage{xparse}
\usepackage{todonotes}
\usepackage{mathtools}
\usepackage{pgfplots}
\pgfplotsset{compat=1.8}
\usepackage[utf8]{inputenc}
\DeclareUnicodeCharacter{00A0}{ }
\usepackage{subcaption}
\definecolor{x11gray}{rgb}{0.75, 0.75, 0.75}
\definecolor{green-yellow}{rgb}{0.68, 1.0, 0.18}
\usepackage{siunitx}
\usepackage{geometry}
\usepackage{graphicx}
\usepackage{booktabs}
\usepackage{array}

\usetikzlibrary{arrows}
\usetikzlibrary{decorations.markings}
\usetikzlibrary{shapes,arrows,chains,calc}
\usetikzlibrary{plotmarks}
\usepgfplotslibrary{patchplots}
\tikzset{%
  >={Latex[width=2mm,length=2mm]},
            base/.style = {rectangle, rounded corners, draw=black,
                           minimum width=2.5cm, minimum height=1cm,
                           text centered, font=\sffamily},
  			Data/.style = {base, fill=green-yellow!15},
     		Process/.style = {base, fill=x11gray!30},
}

\NewDocumentCommand{\tens}{t_}
 {%
  \IfBooleanTF{#1}
   {\tensop}
   {\otimes}%
 }

\usepackage{amssymb,amsmath,amsthm}	
\usepackage{a4wide}
\usepackage{graphicx}      										
\usepackage{url}
\usepackage{dsfont}
\usepackage{xspace}
\usepackage{framed}

\newtheorem{remark}{Remark}[]

\newtheorem{algorithm}{Algorithm}[]

\newcommand{\R}{\ensuremath{\mathbb{R}}}

\begin{document}
\begin{frontmatter}

\title{Simplified ResNet approach for data driven prediction of microstructure-fatigue relationship}

\author[add1]{Christian Gebhardt\corref{cor1}}
\ead{c.gebhardt@iwm.rwth-aachen.de}
\author[add2]{Torsten Trimborn}
\ead{trimborn@igpm.rwth-aachen.de}
\author[add1]{Felix Weber}
\ead{f.weber@iwm.rwth-aachen.de}
\author[add1]{Alexander Bezold}
\ead{a.bezold@iwm.rwth-aachen.de}
\author[add1]{Christoph Broeckmann}
\ead{c.broeckmann@iwm.rwth-aachen.de}
\author[add2]{Michael Herty}
\ead{herty@igpm.rwth-aachen.de}

\address[add1]{Institute for Materials Applications in Mechanical Engineering, RWTH Aachen University}
\address[add2]{Institute für Geometrie und Praktische Mathematik, RWTH Aachen University}

\cortext[cor1]{Corresponding Author}

\begin{abstract}
The heterogeneous microstructure in metallic components results in locally varying fatigue strength. Metal fatigue strongly depends on size and shape of non-metallic inclusions and pores, commonly referred to as "defects". Nodular cast iron (NCI) contains graphite inclusions (nodules) whose shape and frequency influence the fatigue strength. Fatigue strength can be simulated by micromechanical finite element models. The drawback of these models are the large computational costs. Therefore, we employ a data-driven machine learning methodology. More precisely, we utilize the simplified residual neural network (SimResNet) which was recently introduced in \cite{herty2020kinetic} to predict fatigue strength from metallographic data. For the training, we use fatigue data which is simulated with a micromechanical model and the shakedown theorem. The micromechanical models are derived directly from micrographs of nodular cast iron, respectively. The application of SimResNet shows a good performance to predict fatigue strength by local microstructures of nodular cast iron. We show several test cases. The simplified character of SimResNet enables fast predictions of fatigue by microstructures, even in comparision to classical residual neural networks.
\end{abstract}


\end{frontmatter}

\section{Introduction}
Fatigue cracks result from irreversible mechanisms at the microscale of a material. They form at stress levels significantly below the macroscopic yield strength. Multiple fatigue damage mechanisms can operate simultaneously. Subsurface crack initiation depends strongly on the microstructure since their morphology influences stress concentration and localized plastic strain \cite{Castelluccio.}. Thus, metal fatigue strongly depends on size and shape of non-metallic inclusions and pores, commonly referred to as "defects" \cite{Zerbst.2019}. 

Nodular cast iron (NCI) contains graphite inclusions (nodules) whose shape and frequency varies significantly with local cooling conditions and alloy composition \cite{Huebner.2007}. When graphite nodules have a spherical shape, several authors \cite{Endo.1989,Sofue.1979,Niimi.1971} noticed an increase of the fatigue strength with decreasing nodule diameter. However, if the nodules are non-spherical increased cyclic plasticity is observed with Digital Image Correlation (DIC) close to graphite nodules. Strain localization is responsible for the initation of microcracks and is observed in their vicinity \cite{Verdu.2008}. Microcracks may stop growing when their length reaches a size in the dimensions of the initiating defect size. Others continue to grow with short crack behaviour and eventually lead to macroscopic failure.

The strong dependence on the geometry of graphite nodules allows a data-driven prediction of the fatigue strength.
The use of machine learning algorithms has gained a lot of interest in the past decade \cite{jordan2015machine, joshi2019machine, solomonoff2006machine}. Besides the  data science problems like clustering, regression, image recognition or pattern formation there are novel applications in the field of engineering, as e.g.~for production processes \cite{onar2018changing, schmitt2018advances, tercan2017improving}. Also, application of artifical neural networks (ANN) has recently become a trend in material science since ANN models are more flexible than conventional regression models. They provide useful insights in both morphological and mechanical characteristics of a microstructure that have influence over macroscopic mechanical properties. 

For example, Ali et al. \cite{Ali.2019} demonstrated that ANN can significantly decrease the computational cost for a crystal plasticity based framework to model the stress-strain and texture evolution since the ANN model shows up to 99.9 computational speedup over conventional crystal plasticity finite element based models. Moreover, the model showed good predictive capabilities outside the bounds of the original data set.
 
A series of studies adressed prediction of elastic properties by microstructural parameters:
Balokas et al. \cite{Balokas.2018} studied how uncertainties influence the effective elastic properties of 3D braided composites. A Monte Carlo framework was used to generate representative volume elements (RVE). Effective elastic properties were calculated via finite element method (FEM). The data was utilized as samples for an ANN. The ANN can succesfully predict elastic parameters including uncertainties in the yarn volume fraction. Similarily, Shabani et al. \cite{Shabani.2012} studied the influence of volume fraction of $Al_2O_3$ particles on the effective quasi-static properties of an aluminium alloy. Lucon et al. \cite{Lucon.2007} used ANN in place of a conventional micromechanical approach in order to predict macroscopic elastic properties of composite materials. Local microscopic properties and geometry were taken as features for the ANN.

Analysis of microstructure-fatigue relationsip using an ANN was carried out by Geng et al. \cite{Chen.2019}. The data was obtained by applying the shakedown theorem on WC-Co microstructures. Among all features considered, the highly stressed volume fraction of the Co binder-phase showed the best performance of the ANN.

The drawback in application of data-driven models were pointed out by Zhang et al. \cite{Zhang.2003} in a review on polymer composites. Deep neural networks are believed to be able to represent any functional relationship between input and output if enough neurons exist in the hidden layers. This may cause the problem of so-called overfitting \cite{Piotrowski.2013}. The error after training is small due to the powerful ANN learning process. However, when the network is applied to new data the error may rise sharply.

Therefore, in this study, we focus on deep residual neural networks (ResNet) which date back to the 1970s and have been heavily influenced by the pioneering work of Werbos \cite{werbos1994roots}. ResNet have been successfully applied to a variety of applications such as image recognition \cite{wu2019wider}, robotics \cite{zeng2018robotic} or classification \cite{fawaz2018data}. More recently, also applications to mathematical problems in numerical analysis~\cite{HesthavenRay2018,HesthavenRay2019,WangHesthavenRay} and optimal control~\cite{Osher2019} have been studied.

A ResNet can be shortly summarized as follows: Given inputs, which are usually measurements, the ResNet propagates those to a final state. This final state is usually called output and aims to fit a given target. In order to solve this optimization procedure, parameters of the ResNet needs to be optimized and this step is called training. The parameters are distinguished as weights and biases. For the training of ResNets backpropagation algorithms are frequently used \cite{watanabe1990learning, werbos1994roots}.

In this study, we apply a simplified ResNet (SimResNet) \cite{herty2020kinetic} to a micromechanical model for the simulation of fatigue strength of nodular cast iron. In contrast to ANN, the SimResNet is simplified since it utilizes a single neuron with multiple layers per feature. Thus, the above mentioned disadvantages can be controlled much easier. The SimResNet represents the microstructure fatigue relationship well, although only a single micrograph was used for training, respectively. The simplified character of the ANN makes the analysis of influencing shape parameters on the fatigue strength of the investigated material accessible.

In section \ref{Material}, we characterize the microstructure of the cast iron grade investigated with shape parameters. In section \ref{Shakedown}, we present a methodology to built micromechanical finite element models from cast iron micrographs. Furthermore, we introduce the static shakedown theorem which is used to determine a fatigue strength from the microstructure. In section \ref{results}, we explain the simplified ResNet which is used to derive a data-driven microstructure-fatigue relationship. Here, the shape parameters determined in \ref{Material} are used as input features of the SimResNet and the shakedown limits of the micrographs are taken as targets, respectively. The results are discussed and summarized in section \ref{conclusions}.

\section{Metallographic Data}\label{Material}
\begin{figure}[H]
     \centering
     \includegraphics[width=\textwidth]{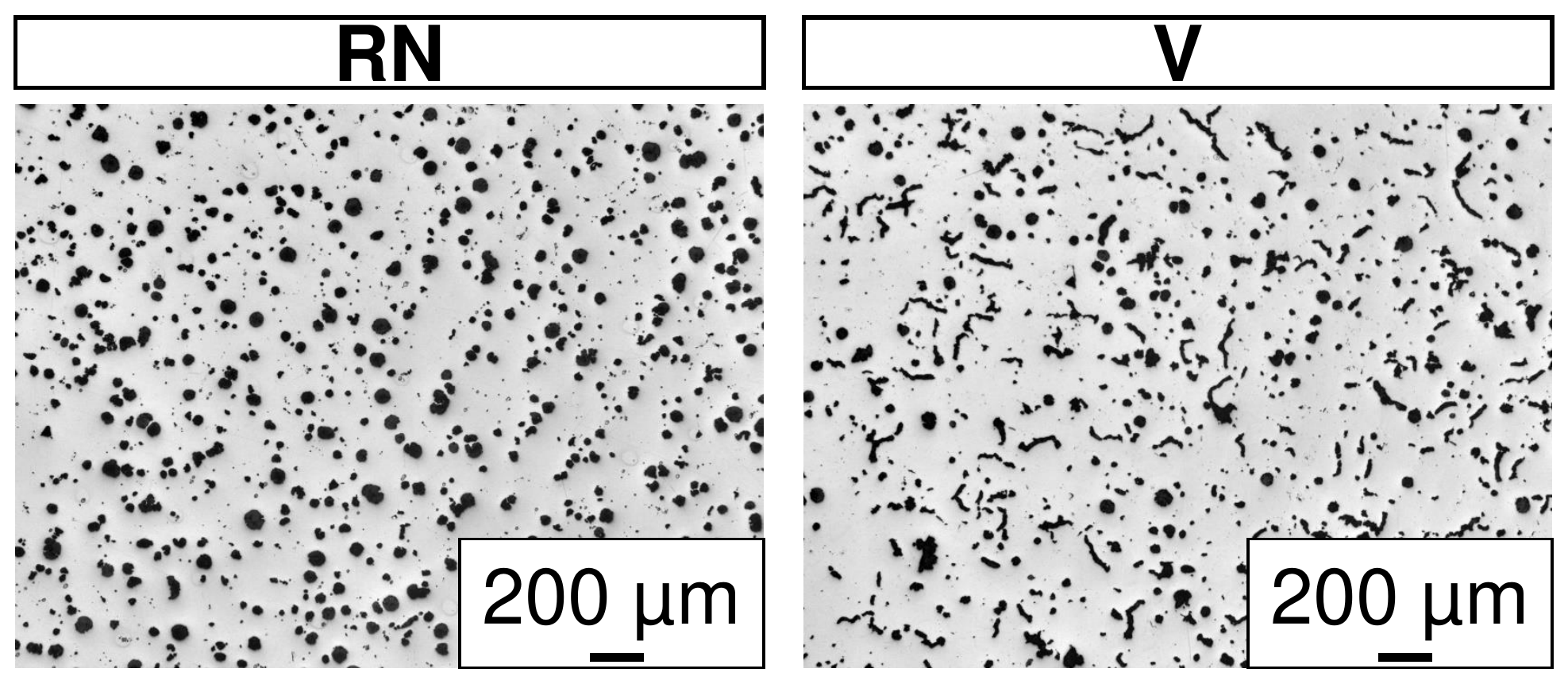}
	\caption{Exemplary micrographs of investigated groups; left: spherical graphite; right: vermicular graphite (V).}
	\label{fig:metallo}
\end{figure}

In this section, we characterize the microstructure of the cast iron grade investigated with shape parameters. For this purpose, metallographic specimens are taken from fatigue samples investigated in \cite{Gebhardt.2018}. The fatigue samples are taken from ingots with the geometry 100 x 100 x 200 mm. The metallographic specimens were ground, polished and investigated with an AxioImager M2m Zeiss light microscope with ProgRes SpeedXTcore5 Jenoptik camera. The micromechanical models (RVE) in section \ref{Shakedown} are built from micrographs of 100x magnification. Subject of the study are two variants of the standardised material EN-GJS-500-14 \cite{DINDeutschesInstitutfurNormunge.V..April2019} with varying graphite morphologies (Fig. \ref{fig:metallo}, Tab. \ref{tab:alloy}).

\begin{table}[!h]
 	\centering
	\caption{Chemical composition of investigated alloys in wt.$\%$.}
	\label{tab:alloy}
	\begin{tabular} {lllll}
		\hline
		\textit{Alloy}&\textit{C}& \textit{Si} & \textit{Ce} & \textit{Mg} \\ 
		\hline
		RN &3.07&3.85&$<$0.0033&0.0280\\
		V &3.07&3.95&$<$0.0033&0.0218\\
		\hline
	\end{tabular}
\end{table}

The microstructure consists of a purely ferritic matrix and graphite, mostly spherical, inclusions called nodules. The graphite morphology was analysed using the open-source software ImageJ Fiji \cite{Schindelin.2012}. The area of a graphite nodule is calculated after binarization utilizing the Otsu's method \cite{Otsu.1979}. The maximum feret's diameter (max. feret) is the longest distance between any two points along the boundary of a graphite nodule. The aspect ratio of the graphite nodule's fitted ellipse is $AR = \frac{Major Axis}{Minor Axis}$.  The results in Tab. \ref{tab:shape_para} are average values per micrograph and show a unique difference between \textit{V} and \textit{RN} regarding the graphite morphology. The data per micrograph was also fitted using a log-normal probability density function (PDF). Results show a larger scatter of the morphological parameters for \textit{V} compared to \textit{RN}.

\begin{table}[h]
	\centering
	\begin{tabular}{lll}
		\hline
		& RN & V \\
		\hline
		Number of microsections [ ] & 97 & 72 \\
		Aspect ratio (avg) [ ] & 0.7049 & 0.67 \\
		Max. feret (avg) [\si{\micro\meter}] & 128 & 190.1 \\
		Area (avg) [\si{\micro\meter\squared}] & 682.23 & 657.67 \\
		\hline
	\end{tabular}
	\caption{Morphological parameters of investigated materials.}
	\label{tab:shape_para}
\end{table}

\begin{figure}[H]
     \centering
     \includegraphics[width=\textwidth]{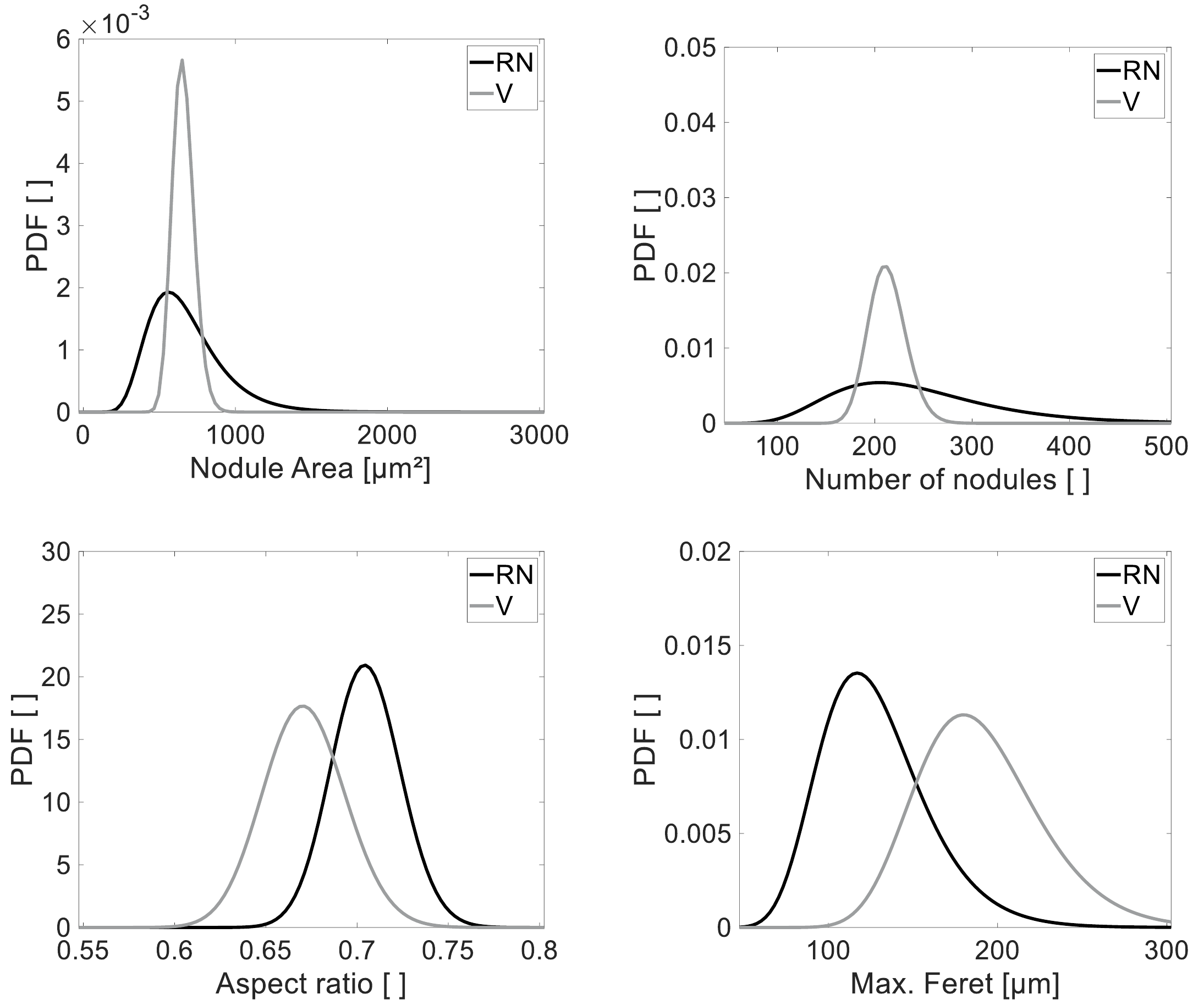}
	\caption{Lognormal fit probability density functions (PDF) for average values out of image analyses for the investigated micrographs of materials \textit{RN} and \textit{V}.}
	\label{fig:distr}
\end{figure} 

\section{Micromechanical Model}\label{Shakedown}
\begin{figure}[H]
     \centering
     \includegraphics[width=\textwidth]{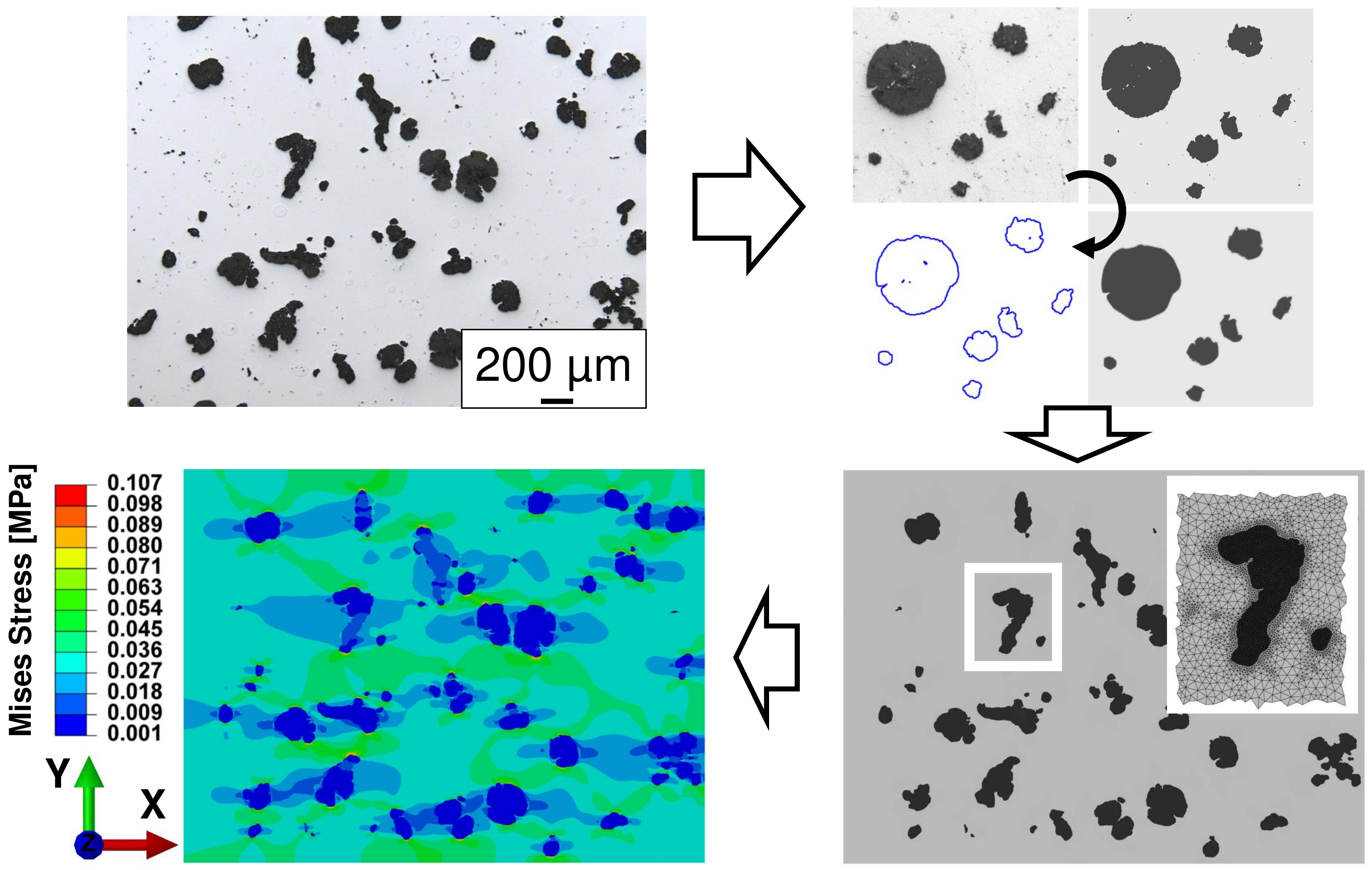}
	\caption{Generation of the linear elastic FE-model.}
	\label{fig:RVE_Gen}
\end{figure}

In this section, we present a methodology to built micromechanical finite element models from cast iron micrographs. Furthermore, we introduce the static shakedown theorem which is used to determine a fatigue strength from the microstructure.
The micromechanical finite element models, called representative volume elements (RVE), are built from light microscopic micrographs using an in-house framework, which is shown in Fig. \ref{fig:RVE_Gen}.

This framework is Python based, utilizing the open packages of sci-kit \cite{Pedegrosa.2011}, sci-kit image \cite{VanderWalt.2014} and open-cv \cite{Bradski.2000}. Thereby, the process of contour extraction is separated into three subsequent steps: First, the micrograph is preprocessed in order to increase the image quality and accessibility. Second, the boundary of graphite inclusions is determined using contour tracing methods. Third, a B-spline curve is constructed using the extracted boundary.

\paragraph*{Contour Extraction}
The micrographs shown in Fig. \ref{fig:metallo} are subject to the influence of changing light conditions and sample preparation. Thus, the preprocessing of the micrograph serves the stabilization of the contour tracing. A normalization is applied initially to ensure contrast quality, since contour detection algorithms are based on contrast detection. A k-means based color clustering is performed on the normalized micrographs, leading to an image containing only a predefined number of colors, in this application one color for the ferrite and graphite phase, respectively. In order to reduce the amount of false artificial pixels, a noise reduction is performed on the clustered image. After noise reduction, the regions occupied by graphite nodules are homogeneous regions of the same color. The image serves as a mask for the original image to place the graphite nodules on a plain white background. This preprocessing ensures the stability of contour extraction algorithms and thus avoids falsely detected contours. Furthermore, residues from sample preparation are removed from the image and thus from further analysis. Using this simplified micrograph, the contour tracing of the graphite nodules is performed by a marching squares algorithm \cite{Lorensen.1987}. Furthermore, the contour points on the boundary of the graphite nodules are placed along a predefined color value leading to an iso-color line. In order to create a mathematical contour description for individual graphite nodules, a non-periodic clamped B-spline is constructed using the contour points directly as control points.

\paragraph*{Meshing}
To solve the shakedown problem, an elastic stress field is required, which is calculated via the finite element method (FEM). High stresses are located along the interfaces of ferrite and graphite phases. Therefore, mesh refinement is mandatory in these areas. In this approach, the local color gradient of the image is used for mesh generation. The rate of color change $\mathbf{G}_{i,j}$ in a HSV (hue, saturation, lightness) picture may be calculated by the spatial derivatives of these values:
\begin{equation}
\mathbf{G}_{i,j} = \begin{pmatrix}
\frac{\partial}{\partial x} \\
\frac{\partial}{\partial y}
\end{pmatrix}  
\tens
\begin{pmatrix}
H_{i,j} & S_{i,j} & V_{i,j}
\end{pmatrix},
\end{equation}
with $i,j$ being pixel IDs. This derivative can be solved numerically by applying a finite difference scheme in the form of
\begin{equation}
\frac{\partial b_{i,j}}{\partial x} = \frac{b_{i+n,j} + b_{i+(n-1),j} + ... + b_{i+1,j} - b_{i-1,j} - ... - b_{i-(n-1),j} - b_{i-n,j}}{2T_n},
\end{equation}
where this scheme spans over $2n+1$ pixels in $x$ direction and $T_n$ is the triangular number. Analogously, the $y$ direction is calculated. The application of the Frobenius norm results in a non-negative scalar for the rate of color change $g_{i,j}$ at a grid node:
\begin{equation}
g_{i,j} = \left\| \mathbf{G}_{i,j} \right\|_F .
\end{equation}
The element size should decrease with increasing $g_{i,j}$ (high rate of change color-wise), so that
\begin{equation}
h_{i,j} = e^{-\gamma g_{i,j}}
\label{eq:mesh_size}
\end{equation}
is proposed for the average mesh size $h_{i,j}$ at a specific position, with $\gamma$ being a scaling factor.
The triangular mesh is generated with the open-source software \textit{Gmsh} \cite{Geuzaine.2009}.

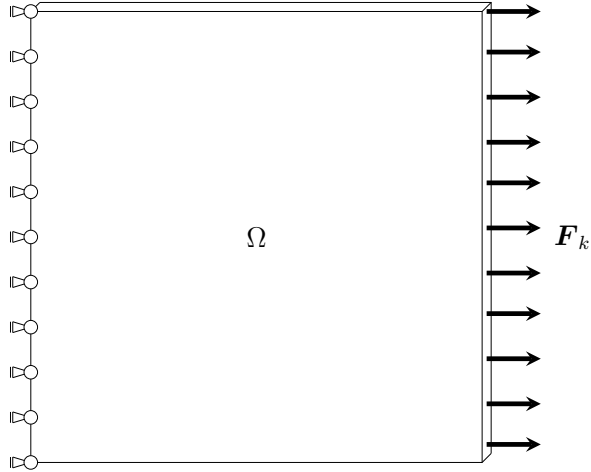
\begin{figure}[h]
	\centering
	\scalebox{0.6}{
		\begin{tikzpicture}
		\draw [black](0,0) -- (10,0) -- (10,10) -- (0,10) -- (0,0);
		\draw [black](10,0) -- (10.2,0.2);
		\draw [black](10,10) -- (10.2,10.2);
		\draw [black](0,10) -- (0.2,10.2);
		\draw [black](0.2,10.2)--(10.2,10.2)-- (10.2,0.2);
		\draw [->,>=stealth, line width=3pt]  (10.1,0.4) -- ++(1.2,0); 
		\draw [->,>=stealth, line width=3pt]  (10.1,1.3) -- ++(1.2,0); 
		\draw [->,>=stealth, line width=3pt] (10.1,2.3) -- ++(1.2,0);
		\draw [->,>=stealth, line width=3pt]  (10.1,3.3) -- ++(1.2,0); 
		\draw [->,>=stealth, line width=3pt]  (10.1,4.2) -- ++(1.2,0);
		\draw [->,>=stealth, line width=3pt]  (10.1,5.2) -- ++(1.2,0); 
		\draw [->,>=stealth, line width=3pt]  (10.1,6.2) -- ++(1.2,0);
		\draw [->,>=stealth, line width=3pt]  (10.1,7.1) -- ++(1.2,0); 
		\draw [->,>=stealth, line width=3pt]  (10.1,8.1) -- ++(1.2,0);
		\draw [->,>=stealth, line width=3pt]  (10.1,9.1) -- ++(1.2,0); 
		\draw [->,>=stealth, line width=3pt] (10.1,10) -- ++(1.2,0); 
		\node [scale=1.5] at (12,5) {\textbf{$\boldsymbol{F}_k$}};
		\node [scale=1.5] at (5,5) {\textbf{$\Omega$}};
		\foreach \y in {0,1,...,10} {
			\draw (0,\y) -- ++(-0.4,-0.12);
			\draw (0,\y) -- ++(-0.4,+0.12);
			\draw (-0.4,\y - 0.12) -- (-0.4,\y +0.12);
			\draw (-0.45,\y - 0.12) -- (-0.45,\y +0.12);
			\filldraw[fill=white] (0,\y) circle (0.15);
			
		}	
		
		\end{tikzpicture}
	}
\caption{Loads and boundary conditions for linear elastic model stress analysis of representative volume element.}
\label{fig:le_model}
\end{figure}

\paragraph*{Linear Elastic Problem}
The shakedown theorem requires the linear elastic stresses $\boldsymbol{\sigma}^E$ as a function of the macroscopic load $\boldsymbol{F}_k$ as input. Fig. \ref{fig:le_model} shows the setup for the simulation of elastic stress with the commerical solver abaqus \cite{abaqus}. 

The local stress field $\boldsymbol{\sigma}$ of the RVE is linked to the macroscopic quantity of the composite over the average stress and strain theorem respectively.
\begin{equation}
\boldsymbol{\Sigma}=\frac{1}{\Omega}\int_{\Omega}\boldsymbol{\sigma}d\Omega .
\end{equation}
Here, the microscopic stress is averaged over the RVE domain $\Omega$, leading to the macroscopic stress $\boldsymbol{\Sigma}$. Consequently, the macroscopic strain is defined as
\begin{equation}
\boldsymbol{E}=\frac{1}{\Omega}\int_{\Omega}\boldsymbol{\varepsilon}d\Omega,
\end{equation}
with the macroscopic strain $\boldsymbol{E}$ and the microscopic strain $\boldsymbol{\varepsilon}$.

\paragraph*{Shakedown Theorem}

In shakedown theory, after some inital cycles of plastic deformation the structure re-enters a state, where it behaves purely elastic in all subsequent cycles. The theory can be applied to micromechanical models \cite{Magoariec.2004}. According to the static shakedown theorem \cite{Melan.1938}, in the shakedown state, the total stress $\boldsymbol{\sigma}$ is divided into a purely elastic stress $\boldsymbol{\sigma}^E$ of a reference body and a time-independent residual stress field  $\bar{\boldsymbol{\rho}}$. Koenig \cite{Konig.2014} has shown that in this case the total plastic dissipation density is bounded. Therefore, no further accumulation of plastic strain occurs if a material enters shakedown state. Furthermore, the normality rule applies and both the total and residual stress must be statically admissible with respect to a convex yield surface, such as the J2-yield surface. With the yield strength defining the yield surface, the material behaviour is elastic perfectly-plastic. The elastic stresses $\boldsymbol{\sigma}^E(\boldsymbol{F}_k)$ in the purely elastic reference RVE are caused by the macroscopic loads $\boldsymbol{{F}_k}$, which span the load domain $\mathcal{L}$. Koenig \cite{Konig.2014} has proved that it is sufficient to consider the vertices of the load polyhedron so defined to determine wether a structure, or in this case the RVE, shakes down for a given combination of macroscopic loads $\left\{ \boldsymbol{F}_k \right\}$.
The shakedown condition can be used as an optimization criterion to find the macroscopic stress that the RVE just shakes down. Enlargement of the load domain defined by $\left\{ \boldsymbol{F}_k \right\}$ is expressed by the safety factor $\alpha$. Therefore, $\alpha \cdot \boldsymbol{\sigma^E}$ gives a lower bound estimation for a macroscopic stress level, for which failure due to localized accumulation of plastic strain does not occur, as growth of microcracks is omitted. The static shakedown theorem writes
\begin{equation} \label{eq:shk}
\begin{aligned}
\underset{\alpha, \bar{\boldsymbol{\rho}}}{\text{maximize}}
&& & \alpha \\
\text{subject to}
&& \nabla \cdot \bar{\boldsymbol{\rho}} &= \boldsymbol{0} \quad in \quad \Omega \\
&& \bar{\boldsymbol{\rho}} \cdot \boldsymbol{n} &= \boldsymbol{0} \quad on \quad \Gamma_1 \\
&& f(\alpha\boldsymbol{\sigma}^E(\boldsymbol{F}_k)+\bar{\boldsymbol{\rho}},\sigma_Y) &\leq 0 \qquad k=1,...,NV. \\
\end{aligned}
\end{equation}
Here, $NV$ denotes the number of load vertices. The static shakedown theorem is expressed in finite element formulation as:
\begin{equation}
\begin{aligned}
\underset{\alpha, \bar{\boldsymbol{\rho}}}{\text{minimize}}
&& & -\alpha \\
\text{subject to}
&& \sum_{i=1}^{NG} \boldsymbol{C}_{i} \bar{\boldsymbol{\rho}}_{i} &= \boldsymbol{0} \\
&& f(\alpha\boldsymbol{\sigma}^E_{i}(F_k)+\bar{\boldsymbol{\rho}}_i)-\sigma_{Y,i} &\leq 0 \\
\end{aligned}
\end{equation}
The index \textit{k} denotes the load vertices and \textit{i} the Gaussian points. The problem is reformulated according to \cite{Chen.2016} to reduce computational efforts.

\begin{figure}[H]
     \centering
     \includegraphics[width=0.75\textwidth]{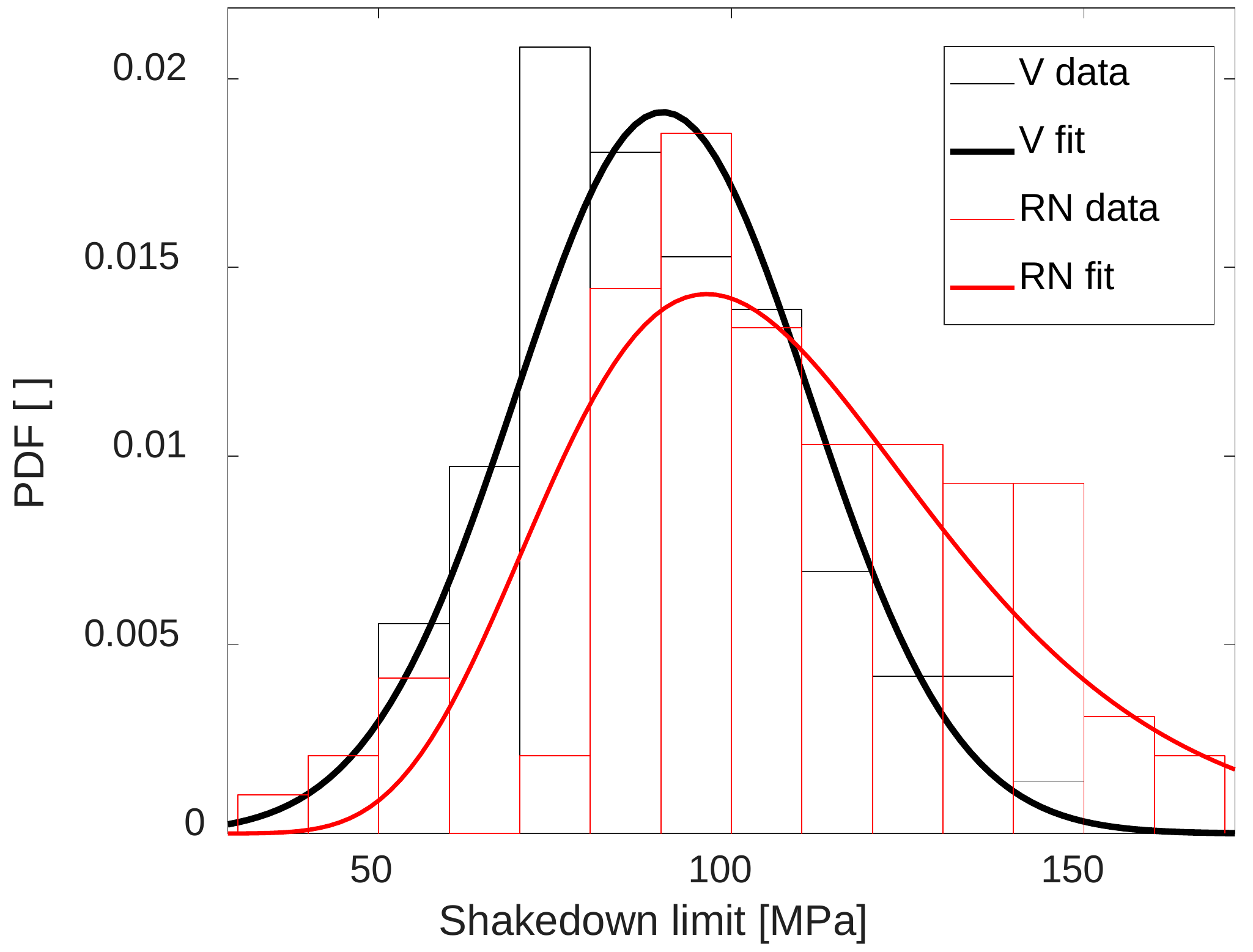}
	\caption{Raw data fitted by log-normal probability density function (PDF) for simulated shakedown limits \textit{RN} and \textit{V}.}
	\label{fig:distr_shk}
\end{figure}

\paragraph*{Simulation Results}
The described method was applied to the micrographs of the groups V and RN (Tab.\ref{tab:shape_para}), respectively. During image processing, graphite nodules with an area smaller than \SI{30}{\micro\meter} were removed from the RVE. Two load vertices were considered in the optimization problem corresponding to a uniaxial pulsating load. The elastic stresses were calculated by the commercial FE-solver Abaqus \cite{abaqus}. The commercial optimizer gurobi was used to solve the optimization problem \cite{gurobi}. The elastic and plastic constants were taken from \cite{Gebhardt.2018}. Output of the simulation is a macroscopic load per micrograph where no failure due to accumulating plasticity occurs. Fig. \ref{fig:distr_shk} shows the raw data and a logarithmic normal distribution fit. Group RN shows a higher mean shakedown limit than group V. However, the scatter of the data is higher.

\begin{table}[h]
	\centering
	\begin{tabular}{lll}
		\hline
		& RN & V \\
		\hline
		Mean & 108.4 & 90.3 \\
		Variance & 945.6 & 441.4 \\
		$\mu$ & 4.64 & 4.48 \\
		$\sigma$ & 0.28 & 0.23 \\
		\hline
	\end{tabular}
	\caption{Parameters of the lognormal distribution in Fig. \ref{fig:distr_shk}.}
	\label{tab:shk_res}
\end{table}


\section{Simplified Residual Neural Network}\label{SimResNet}
\begin{figure}[h!]
\includegraphics[scale=1]{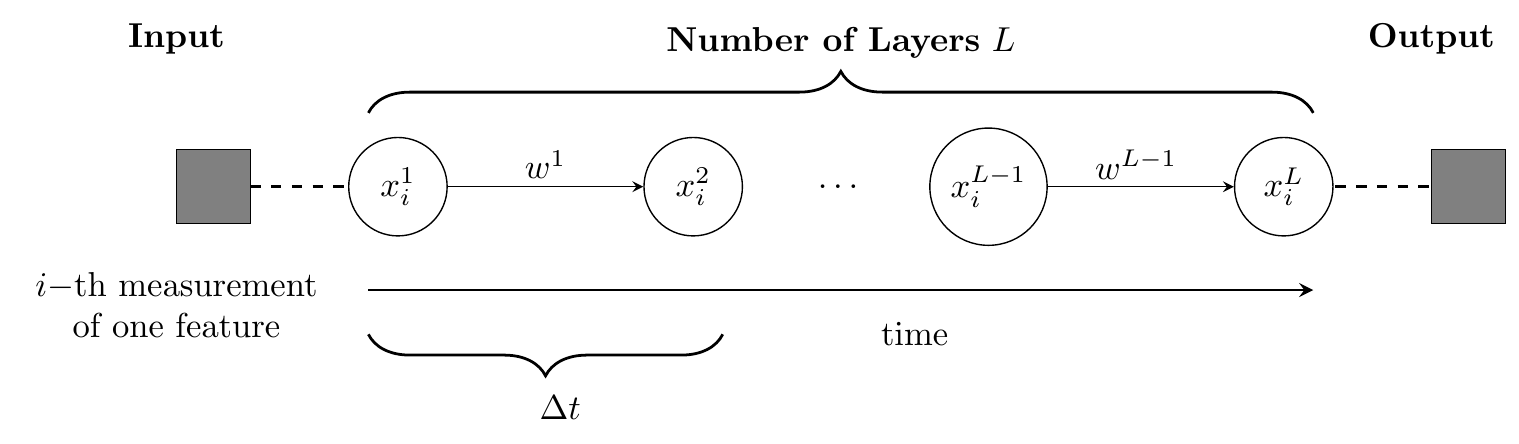}
\caption{Schematic picture of the SimResNet model, where $x_i^k:=x_i(t_k)$ and $w^k:=w(t_k)$ for $k=1,...,L$. }\label{NN}
\end{figure}

In this section, we present the simplified residual neural network (SimResNet) approach recently introduced by Herty et al. \cite{herty2020kinetic}.
We assume that the input signal consists of $d$ features. In our application features are morphological parameters (area, maximum feret and aspect ratio of a graphite nodule).
For simplicity, we assume that the value of each feature is one dimensional and thus the input signals are given by $\boldsymbol{x}_i^0\in\R^d,\ i=1,...,M$. Here, $M$ denotes the number of measurements or input signals. In the following, we assume that the number of neurons is identical in each layer and the number of neurons $N$ in each layer is identical to the number of features $d$. 
This is a huge simplification in comparison to standard ResNet models. Furthermore, we consider $L$ layers and interpret these layers as discretization points. We refer to Figure \ref{NN} for a visualization of the neural network structure. Thus, we define a discretization $t_k:=k\ \Delta t,\ \Delta t>0,\ k=0,...,L$.  The microscopic model which defines the evolution of the activation energy of each neuron $\boldsymbol{x}_i(t)\in\R^d$, with fixed input signal $\boldsymbol{x}_i^0\in\R^d$, weight $\boldsymbol{w}\in\R^{d\times d}$ and bias $\boldsymbol{b}\in\R^{d}$ reads:
\begin{equation} \label{eq:defNN}
 \begin{cases}
 \boldsymbol{x}_i(t_{k+1})=\boldsymbol{x}_i(t_k)+\Delta t\ \sigma\left( \frac{1}{d } \boldsymbol{w}(t_k)\ \boldsymbol{x}_i(t_k)+\boldsymbol{b}(t_k) \right),\\
 \boldsymbol{x}_i(t_0)=\boldsymbol{x}_i^0
 \end{cases}
\end{equation}
for each fixed $i=1,\dots,M$.
Here, $\sigma: \R\to \R$ denotes the activation function which is applied component wise. Frequently utilized examples for the activation function are  given by the  ReLU function $\sigma_R(x) = \max\{0,x\}$ or the sigmoid function $\sigma_S(x) = \frac{1}{1+\exp(-x)}$.

\begin{remark}
In~\eqref{eq:defNN} we have reformulated a neural network by introducing a parameter $\Delta t$ and a time discrete concept which corresponds to the layer discretization. More precisely the time step $\Delta t>0$ may be defined as $\Delta t:= \frac{1}{L}$ and, in this way, we can see~\eqref{eq:defNN} as an explicit Euler discretization of an underlying time continuous model on $t\in[0,1]$. 
A similar modeling approach with respect to the time continuous interpretation of layers has been introduced in \cite{chen2018neural, ruthotto2018deep}. In  the continuum limit which corresponds to $\Delta t\to 0$ and $L\to \infty$ equation \eqref{eq:defNN} becomes:
\begin{equation} \label{eq:continuousNN}
 \begin{cases}
 \dot{\boldsymbol{x}}_i(t) = \sigma\left( \frac{1}{d} \boldsymbol{w}(t)\ \boldsymbol{x}_i(t)+\boldsymbol{b}(t)\right),\\
 \boldsymbol{x}_i(0)=\boldsymbol{x}_i^0,
 \end{cases}
\end{equation}
for each fixed $i=1,\dots,M$.
\end{remark}

\paragraph{Training of NN}
The crucial part in applying a neural network is to train the network. By training one aims to minimize the distance of the output of the neural network
at some fixed time  $T>0$ to the target $\boldsymbol{h}_i\in\R^d$. Mathematically speaking one aims to minimize the distance 
$$
\min\limits_{\boldsymbol{w}, \boldsymbol{b}}|| \boldsymbol{x}_i(T)- \boldsymbol{h}_i||_2^2,
$$
where we use the squared $L^2$ distance between the target and the output to define the loss function. Other choices are certainly possible~\cite{janocha2017loss}.
The procedure can be computationally expensive on the given training set. Most famous examples of such an optimization are so called back propagation algorithms or ensemble Kalman filters \cite{haber2018never, kovachki2019ensemble, watanabe1990learning}.
In this paper we consider the back propagation algorithm as defined in \cite{watanabe1990learning}, which reads as follows :

\begin{algorithm} The \textbf{backpropagation algorithm} can be summarizes as:
\begin{enumerate}
\item $\boldsymbol{w}(t_k)= \boldsymbol{w}(t_k)- \eta\ \Phi(t_k)\ \boldsymbol{x}_i(t_{k-1})$
\item $\Phi(t_L):= (\boldsymbol{x}(t_L)-\boldsymbol{h}_i)\ \sigma'\left( \frac{1}{d } \boldsymbol{w}(t_L)\ \boldsymbol{x}_i(t_L)+\boldsymbol{b}(t_L) \right)$
\item $\Phi(t_k):= \sigma'\left( \frac{1}{d } \boldsymbol{w}(t_L)\ \boldsymbol{x}_i(t_L)+\boldsymbol{b}(t_L) \right)\ \textbf{w}(t_{k+1}) \Phi(t_{k+1}),\ k=L-1,...,1 $
\end{enumerate}
In a similar fashion the bias $\textbf{b}$ can be computed. For details we refer to \cite{watanabe1990learning}.
\end{algorithm}

We reduce overfitting by constraining model complexity. We control the number of layers by the performance of the SimResNet on a validation set. This is known as structural stabilization and has been performed in \cite{cunningham2000stability} as well. As alternative one may also add a regularization term to the objective and to control the weights and biases \cite{zaremba2014recurrent}. 

\paragraph{Advantages of SimResNet} The advantages of SimResNet in comparison to classical artificial neural networs are twofold. First of all due to the reduction of complexity, caused by a fixed and small number of neurons, there is a speed up in the training procedure.  We refer to the Figure \ref{speed} in section \ref{results}. Secondly the SimResNet approach makes it possible to derive partial differential equations. This has been performed in \cite{herty2020kinetic} which gives a probabilistic description of the neuron energy in the case of infinitely many measurements. The authors show that such a kinetic description is applicable to classical machine learning applications. The real achievement of such a kinetic translation is the possibility to analyze the neural network. More precisely, the authors Herty et al. \cite{herty2020kinetic} show that it is possible to derive a priori bounds on the needed simulation time and are able to characterize the optimal weights and bias of the SimResNet model.


\section{Results}\label{results}
Before we present the results of the SimResNet model, we present the data structure of our problem and  state the settings of the SimResNet method in detail.
We consider a one dimensional target which is our shakedown limit. This limit is fixed for each pictures in both data sets (V=non-spherical graphite nodules or RN=spherical graphite nodules). In each picture, we consider morphological parameters of the graphite nodules. 
These parameters are maximum feret, area and aspect ratio. The schematic overview of the data is given in Tab. \ref{data}.

\begin{table}[h!]
\centering
\begin{tabular}{| l | c | c | c | c | c|}
  \hline			
  Picture & Graphite nodule & Max. Feret  & Area & AR & Shakedown Limit\\
  \hline
  \hline
  $1$ & $1$ & $2095$ & $2895$  & $1000$ & $76.86$\\
  \hline
     \vdots & \vdots & \vdots & \vdots  & \vdots & \vdots\\
  \hline
  1 & M & $4189$ & $2288$ & $1141 $& $ 76.86$ \\
  \hline  
    2 & $1$ & $43989$ &$ 1447$ & $1052$ & $97.60$ \\
  \hline
    \vdots & \vdots & \vdots & \vdots  & \vdots & \vdots\\
    \hline
\end{tabular}
\caption{Schematic illustration of the data.}
\label{data}
\end{table}

\begin{figure}[h!]
\begin{center}
\includegraphics[width=0.75\textwidth]{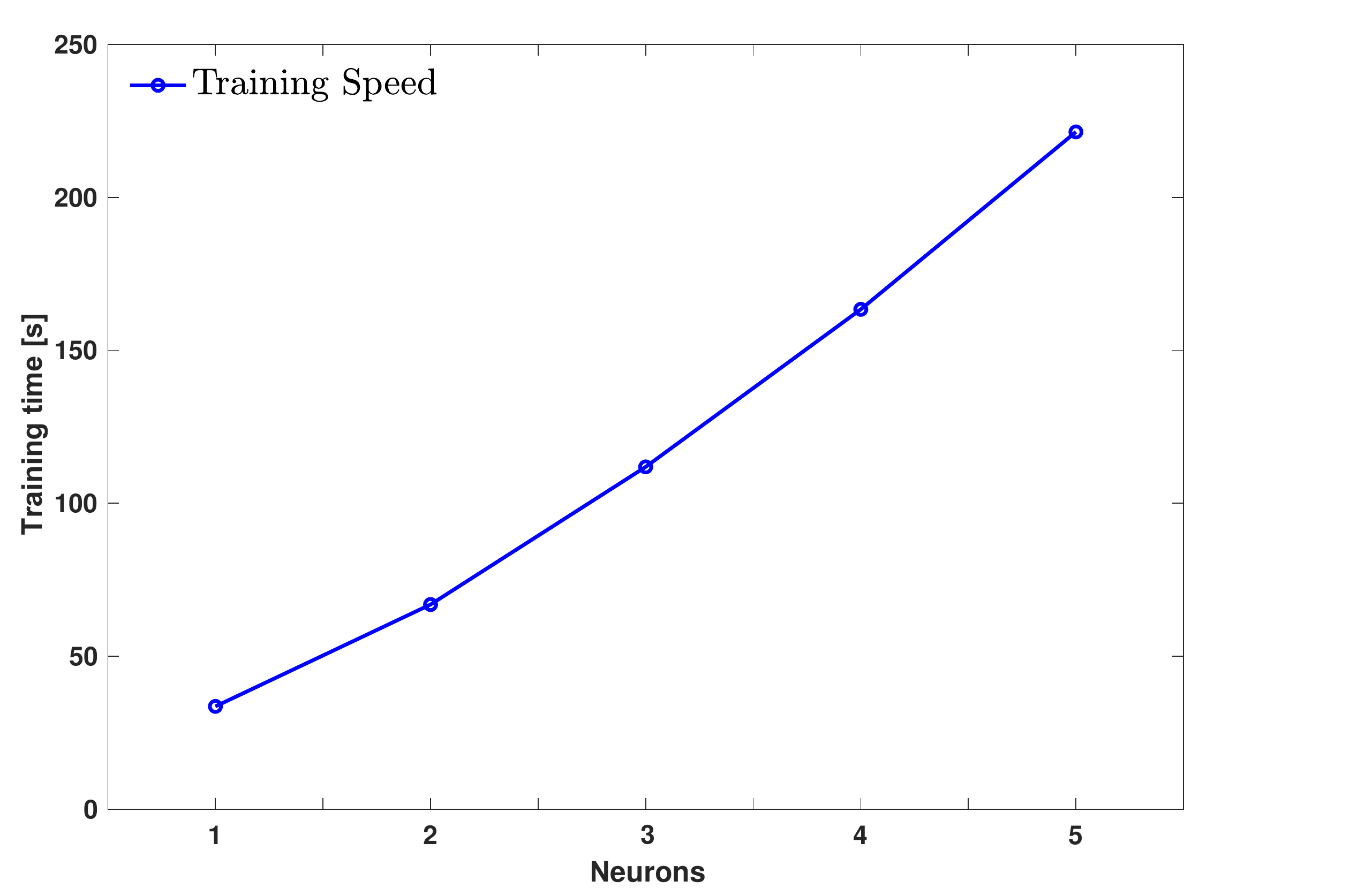}
\end{center}
\caption{Training speed of ResNet for different number of neurons for each feature. The one neuron case ($N=1$) corresponds to the SimResNet model. The number of layers and the training set are identical and fixed.}\label{speed}
\end{figure}

\paragraph{SimResNet Setting}
We consider the sigmoid activation function and four layers, which corresponds to two hidden layers. For the back propagation algorithm we have chosen a learning rate of $\xi=0.1$ and a maximum number of $10^4$ iterations of the training process. This choice has been selected with the help of a validation set in order to avoid overfitting. We have normalized our input and target data for a more comprehensive data vizualization. In the simplest setting we train and validate our SimResNet with the data of one randomly selected picture, respectively.  For our training we always use $M=150$ measurements of each feature. In a second setting we train the SimResNet seperately with $5$ randomly selected pictures separately and compute the average of the computed weights an biases. Tests with more pictures did not improve the results and are therefore omitted. In order to measure the quality of our trained neural network we consider the following error sum for the j-th picture:
$$
\eta_j:= \sum\limits_{i=1}^{M} |\boldsymbol{x}_i(T)-\boldsymbol{h}_j|.
$$
In order to judge the performance over the whole data set of $P$ pictures we consider the sample average $\bar{\eta}$ and  variance  $\theta$ of all errors 
\begin{align*}
&\bar{\eta}:= \frac{1}{P} \sum\limits_{j=1}^P \eta_j,
& \theta:=  \frac{1}{P} \sum\limits_{j=1}^P(\eta_j-\bar{\eta})^2.
\end{align*}
For our simulations, we always consider $P=70$ pictures of each group \textit{V} or \textit{RN}. Figure \ref{speed} depicts the time advantage of SimResNet in comparison to ResNet models with more than one neuron for each feature.

\subsection{Single Feature}
\begin{figure}[h!]
\begin{center}
\includegraphics[width=0.5\textwidth]{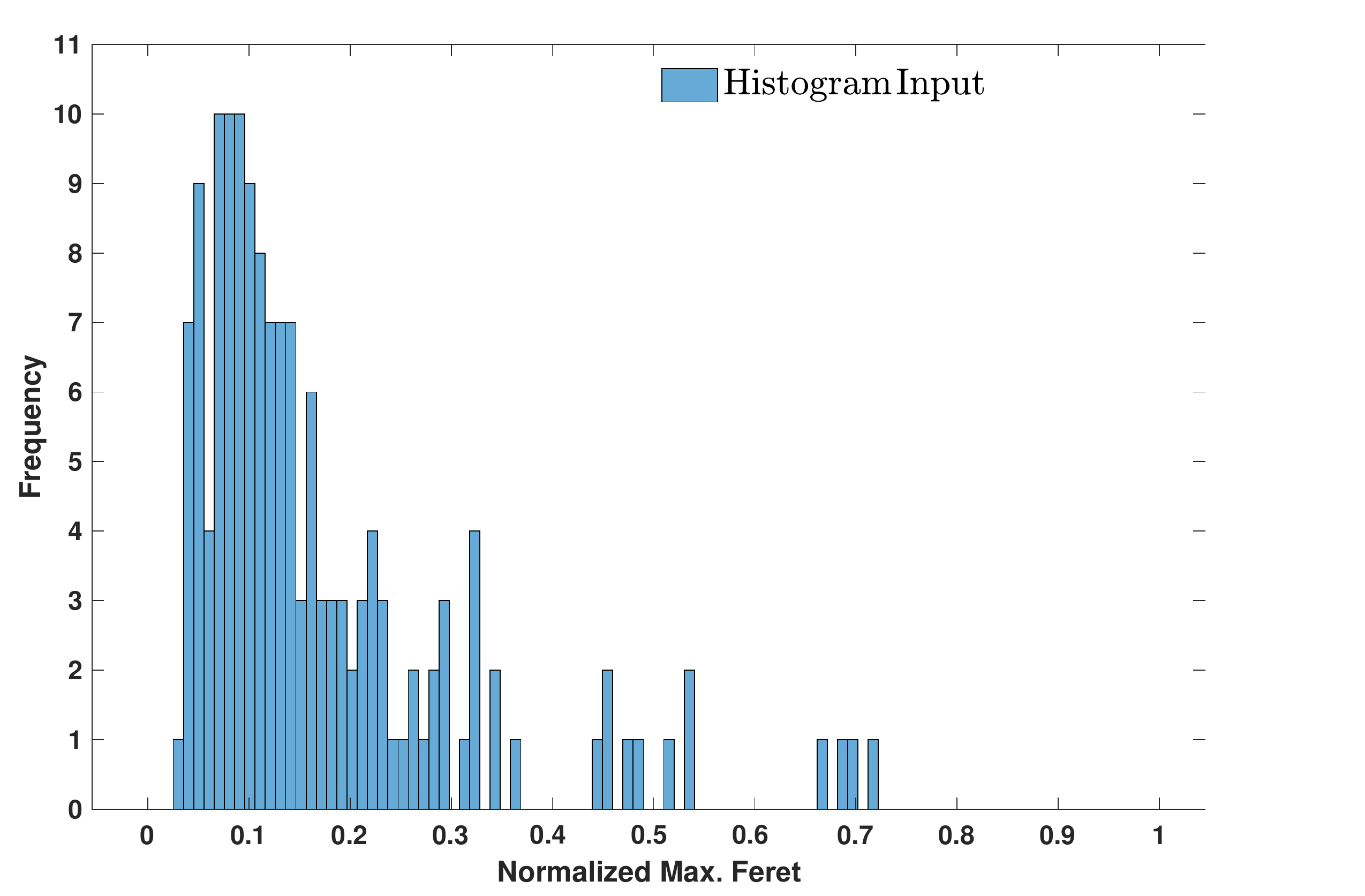}\hfill
\includegraphics[width=0.5\textwidth]{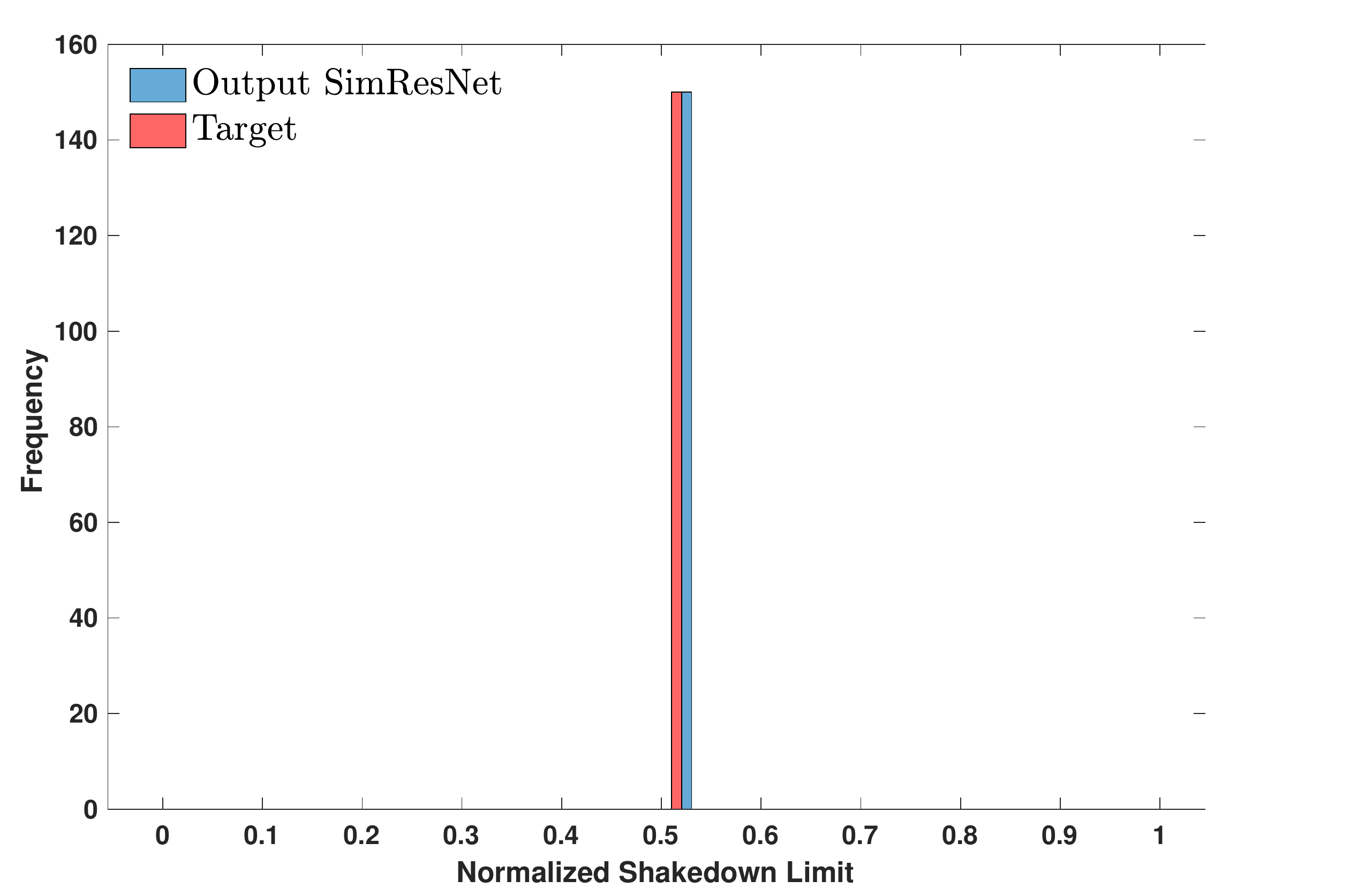}
\caption{Trained SimResNet with one picture on data set V ($\eta_j=1.12$). LHS: Histogram of input maximum feret of data set V. RHS: Output of SimResNet and target. } \label{FeretGood}
\end{center}
\end{figure}

\begin{figure}[h!]
\begin{center}
\includegraphics[width=0.5\textwidth]{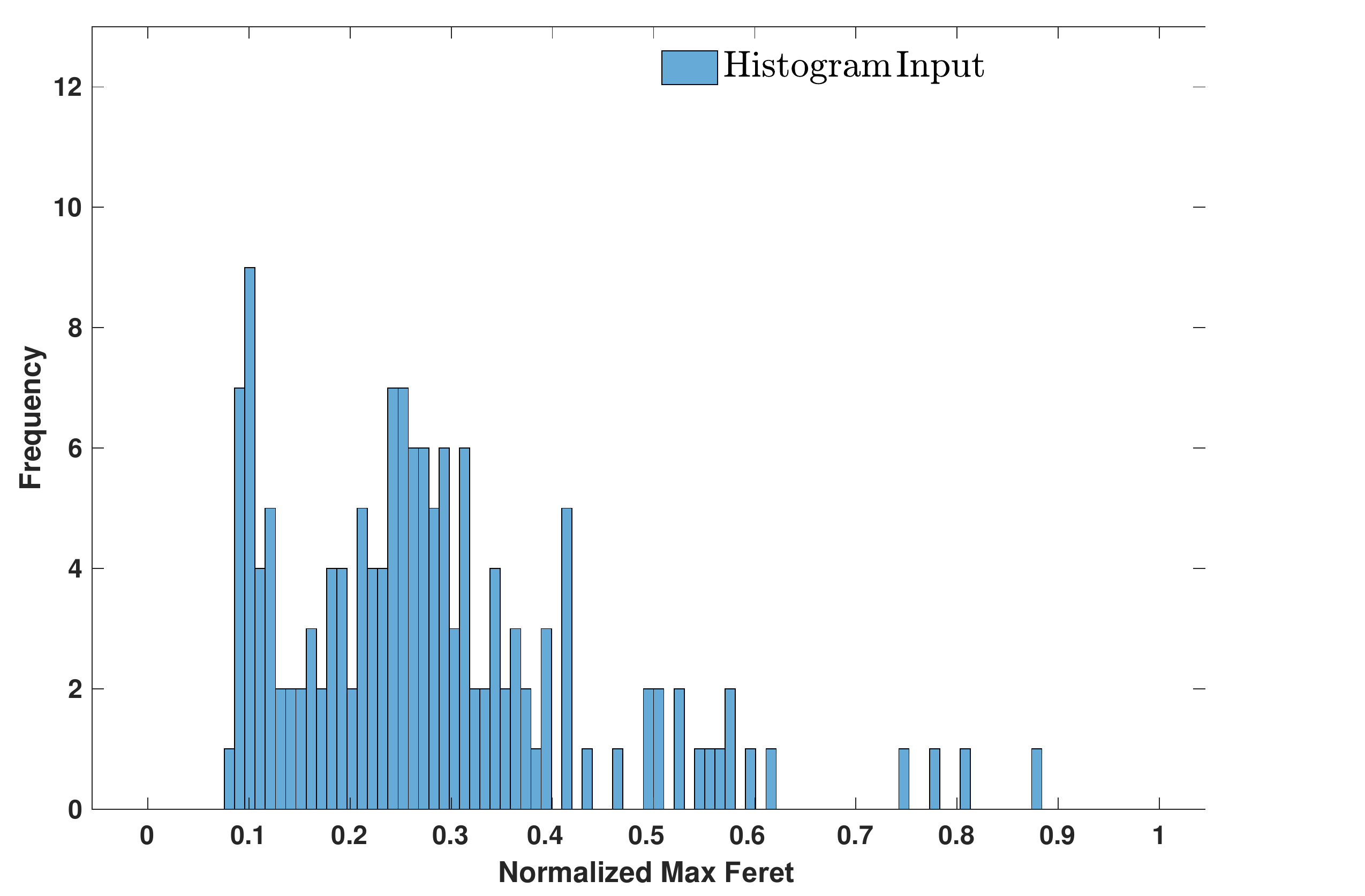}\hfill
\includegraphics[width=0.5\textwidth]{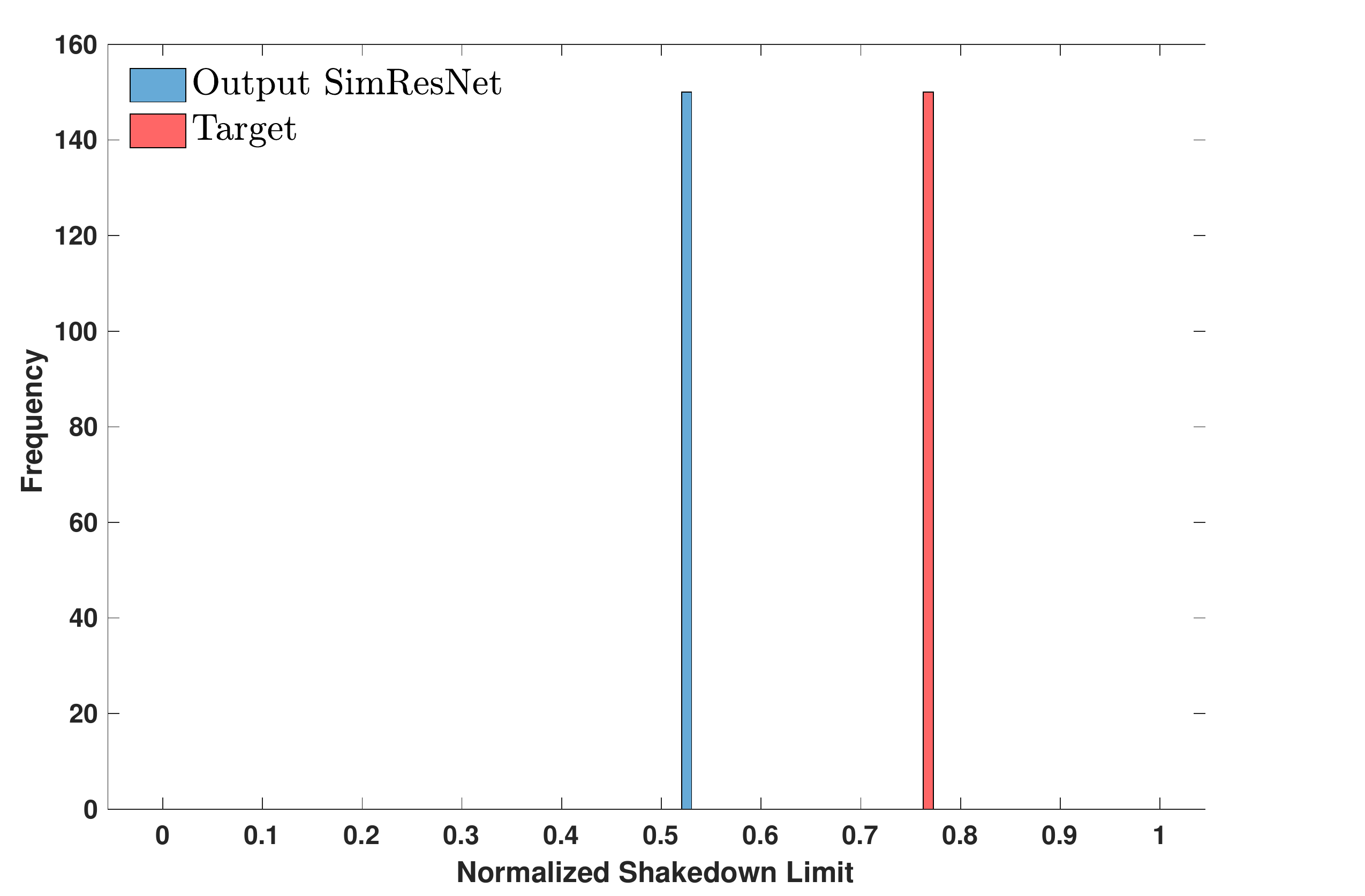}
\caption{Trained SimResNet with one picture of dataset V applied on picture of dataset RN ($\eta_j=35.80$). LHS: Histogram of input maximum feret of data set RN. RHS: Output of SimResNet and target.}\label{FeretBad}
\end{center}
\end{figure}

\begin{figure}[h!]
\begin{center}
\includegraphics[width=0.5\textwidth]{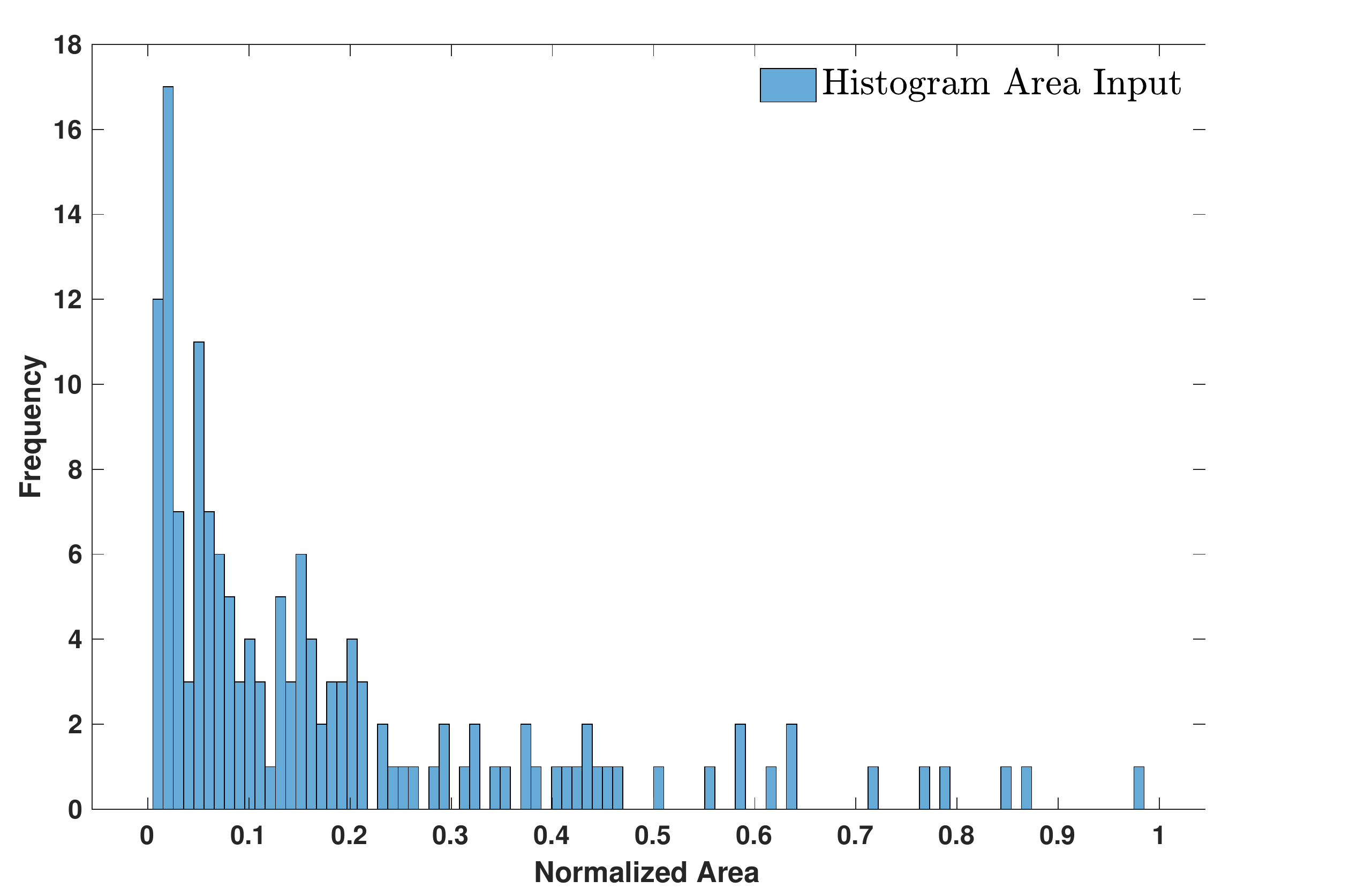}\hfill
\includegraphics[width=0.5\textwidth]{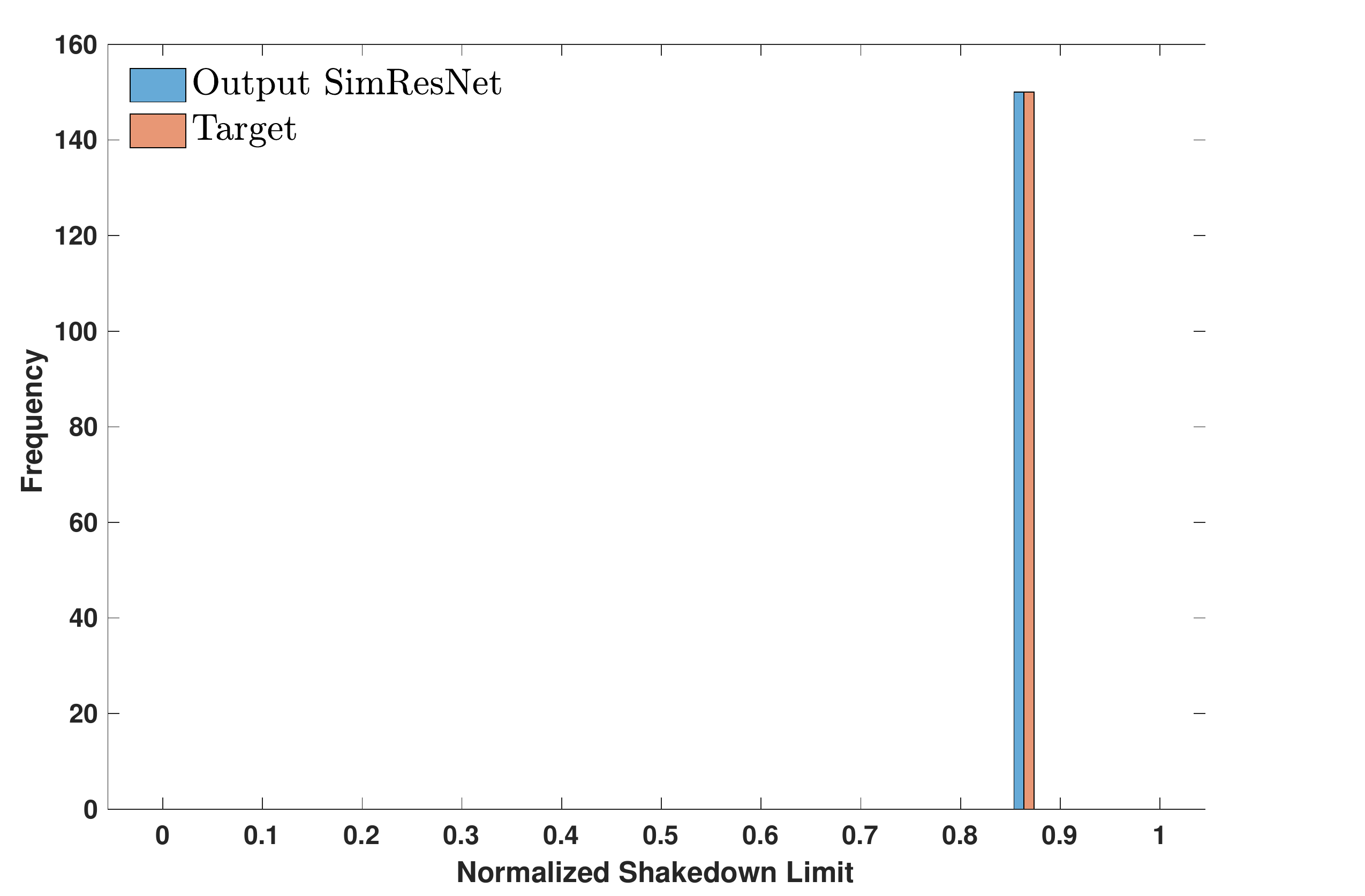}
\caption{Trained SimResNet with one picture on data set RN ($\eta_j=2.26$). LHS: Histogram of input Area of data set RN. RHS: Output of SimResNet and target.}\label{AreaGood}
\end{center}
\end{figure}

\begin{figure}[h!]
\begin{center}
\includegraphics[width=0.5\textwidth]{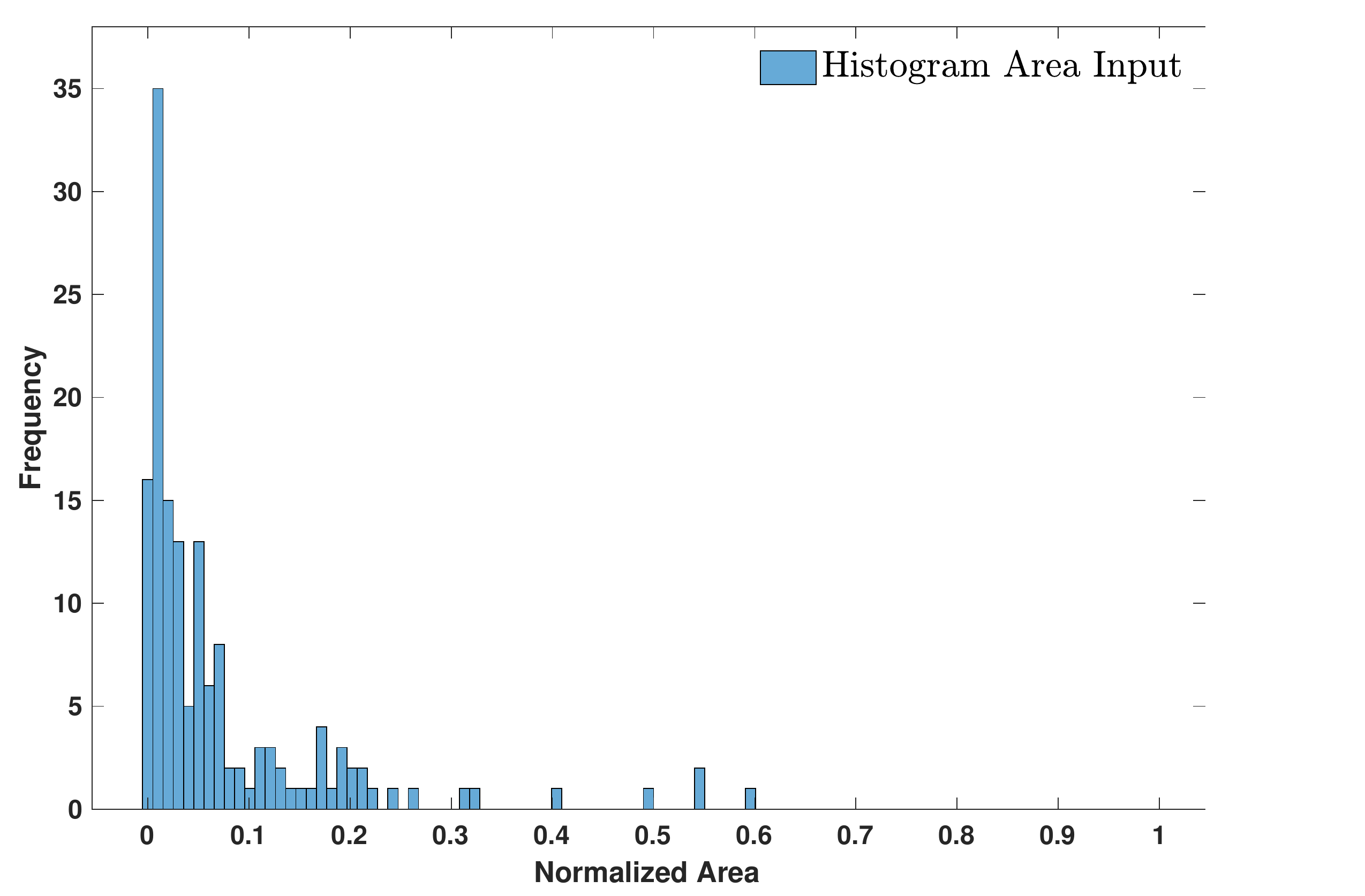}\hfill
\includegraphics[width=0.5\textwidth]{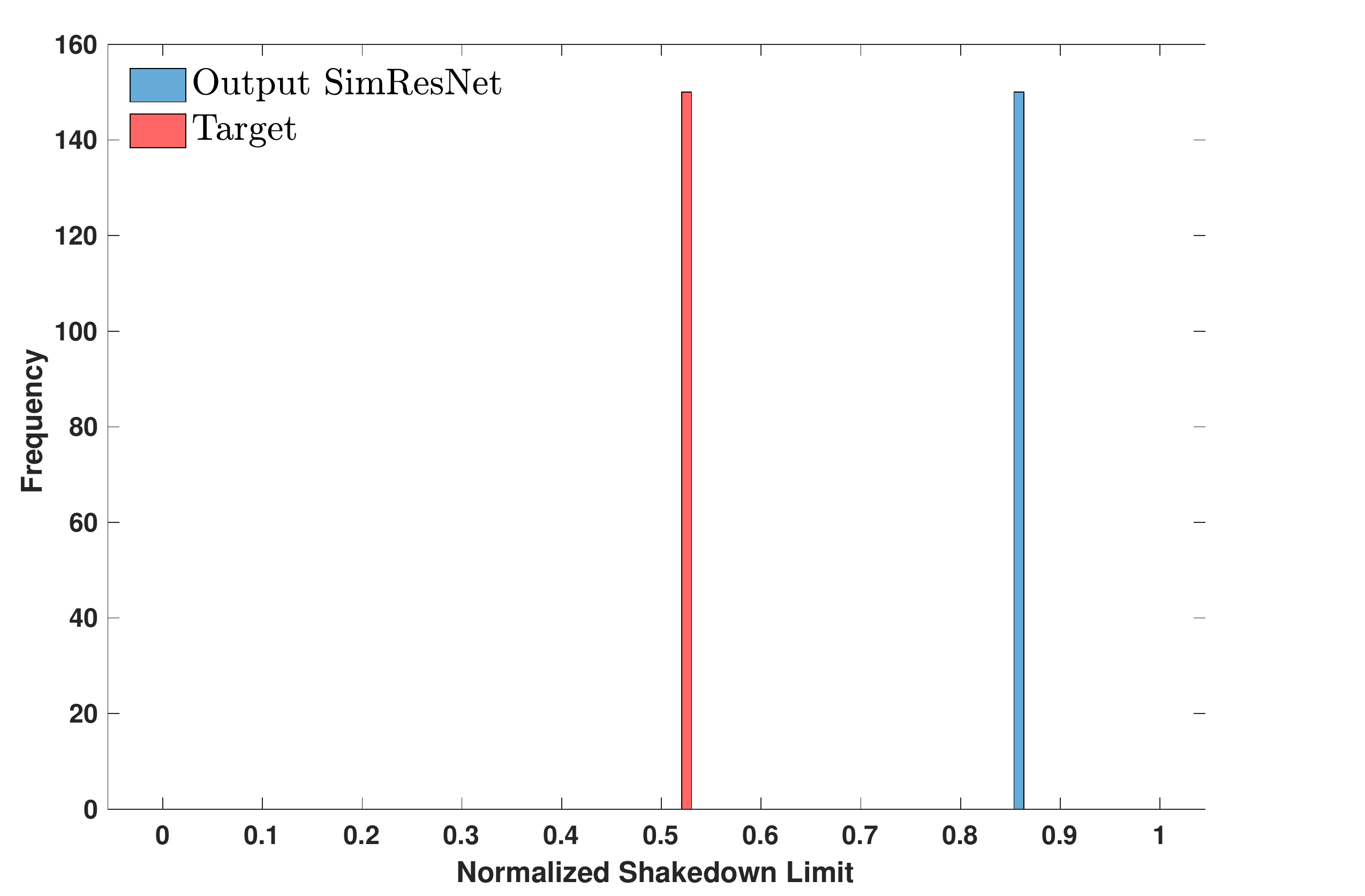}
\caption{Trained SimResNet with one picture of data set RN applied on picture of dataset V ($\eta_j=51.73$). LHS: Histogram of input Area of data set V. RHS: Output of SimResNet and target. }\label{AreaBad}
\end{center}
\end{figure}

We first study the case of one feature. Thus, we consider as input maximum feret, area and aspect ratio (AR) separately. In order to visualize the input data, we use histogram plots. For example, the input maximum feret is depicted in Figures \ref{FeretGood} and \ref{FeretBad} (left hand side). The input area is shown in Figures \ref{AreaGood} and \ref{AreaBad} (left hand side). The training output is shown in the Figures \ref{FeretGood}-\ref{AreaBad} on the right hand side. The target is given as a red bar. The Figures \ref{FeretGood} and \ref{AreaGood} show a good fit of the output of the SimResNet to the desired target. 
In comparison to that we obtain a bad fit in the cases where a trained SimResNet of the data set V (RN) is applied on the data set RN (V) (see Figures \ref{FeretBad}, \ref{AreaBad}).
The reason can be obtained in the different shapes of the histograms of the input on each data set  which can be clearly deduced in the Fig. \ref{FeretGood}-\ref{AreaBad}.  

A comprehensive overview of the results of our SimResNet with respect to the error measures $\bar{\eta}$ and $\theta$ are given in Tables \ref{OneFeature1} and \ref{OneFeature5}. 
First we notice that the errors are low for both data sets and are of the same order. Interestingly are the errors for each different input (maximum feret, area, AR) comparable for both data sets V and RN.
Furthermore, we notice that the errors and the variance of the errors are significantly lower in the V data set than in RN. 
This implicates that the input distributions (histograms) are more homogeneous in the data set V than in RN. In addition, Tables \ref{OneFeature1} and \ref{OneFeature5} reveal that an improved training procedure leads only to a slight improvement of the results. 

\begin{table}[h!]
\centering
\begin{tabular}{| l | c | c | c | }
  \hline			
  Data Set & Max. Feret  & Area & AR\\
  \hline
  \hline
  V & $ \bar{\eta}=1.55,\quad \theta= 1.74$  & $\bar{\eta}=1.51,\quad \theta=1.74$  & $\bar{\eta}=1.56 ,\quad \theta=1.86$ \\
  \hline
  RN & $\bar{\eta}= 2.37,\quad \theta=2.99$ &$ \bar{\eta}= 2.35,\quad \theta= 3.02$& $\bar{\eta}=2.27 ,\quad \theta= 2.82$ \\
  \hline  
\end{tabular}
\caption{Error quantities for each feature trained with one picture.}
\label{OneFeature1}
\end{table}

\begin{table}[h!]
\centering
\begin{tabular}{| l | c | c | c | }
  \hline			
  Data Set & Max. Feret  & Area & AR\\
  \hline
  \hline
  V & $ \bar{\eta}=1.45,\quad \theta=1.70 $  & $\bar{\eta}=1.49,\quad \theta=1.75$  & $\bar{\eta}=1.44 ,\quad \theta=1.63$ \\
  \hline
  RN & $\bar{\eta}= 2.17,\quad \theta=3.12$ &$ \bar{\eta}=2.35 ,\quad \theta=3.02 $& $\bar{\eta}=2.27 ,\quad \theta=2.82 $ \\
  \hline  
\end{tabular}
\caption{Error quantities for each feature trained with five pictures.}
\label{OneFeature5}
\end{table}

\clearpage

\subsection{Multiple Features}
In the case of multiple features, the results are comparable to the one feature case. 
Nevertheless, we obtain small deviations. In the two feature case, there is a minor improvement for the RN data set in comparison to the one feature case (see Table \ref{twoF1}). 
As in the one feature case, we obtain an improvement in the case of a larger data set (compare Tables \ref{twoF1} and \ref{twoF5}). In the three feature case, a similar reduction can be obtained for the different training procedures (compare Tables \ref{threeF1} and \ref{threeF5}). The error quantities in the three feature case are of a similar magnitude as in the other settings. The best result is obtained in the three feature case with improved training. However, the error quantities do not improve significantly.

\begin{table}[h!]
\centering
\begin{tabular}{| l | c | c | c | }
  \hline			
  Data Set & Max. Feret + Area & Max. Feret + AR& AR + Area \\
  \hline
  \hline
  V & $\bar{\eta}=1.55,\quad \theta=1.73$ & $\bar{\eta}=1.55,\quad \theta=1.74$ & $\bar{\eta}=1.56,\quad \theta=1.73$ \\
  \hline
  RN & $\bar{\eta}=2.34 ,\quad \theta =3.05 $ & $\bar{\eta}=2.37,\quad \theta=3.01$ & $\bar{\eta}=2.37,\quad \theta=3.02$ \\
  \hline  
\end{tabular}
\caption{Error quantities for the two feature case, trained with one picture.}
\label{twoF1}
\end{table}

\begin{table}[h!]
\centering
\begin{tabular}{| l | c | c | c | }
  \hline			
  Data Set & Max. Feret + Area & Max. Feret + AR& AR + Area \\
  \hline
  \hline
  V & $\bar{\eta}=1.48,\quad \theta=1.64$ & $\bar{\eta}=1.61,\quad \theta=1.88$ & $\bar{\eta}=1.63,\quad \theta=1.91$ \\
  \hline
  RN & $\bar{\eta}=2.32 ,\quad \theta =3.01	 $ & $\bar{\eta}=2.28,\quad \theta=2.95$ & $\bar{\eta}=,2.27\quad \theta=2.93$ \\
  \hline  
\end{tabular}
\caption{Error quantities for the two feature case, trained with five picture.}
\label{twoF5}
\end{table}

\begin{table}[h!]
\centering
\begin{tabular}{| l | c |  }
  \hline			
  Data Set & Max. Feret + Area +AR \\
  \hline
  \hline
  V & $\bar{\eta}=1.53,\quad \theta=1.73$  \\
  \hline
  RN & $\bar{\eta}=2.34 ,\quad \theta =3.05 $ \\
  \hline  
\end{tabular}
\caption{Error quantities for the three feature case, trained with one picture.}
\label{threeF1}
\end{table}

\begin{table}[h!]
\centering
\begin{tabular}{| l | c |  }
  \hline			
  Data Set & Max. Feret + Area + AR \\
  \hline
  \hline
  V & $\bar{\eta}=1.46,\quad \theta=1.56$  \\
  \hline
  RN & $\bar{\eta}=2.32 ,\quad \theta =3.02 $ \\
  \hline  
\end{tabular}
\caption{Error quantities for the three feature case, trained with five picture.}
\label{threeF5}
\end{table}

\subsection{Comparison of Simulated and Predicted Data}
\begin{figure}[h!]
\begin{center}
\includegraphics[width=0.75\textwidth]{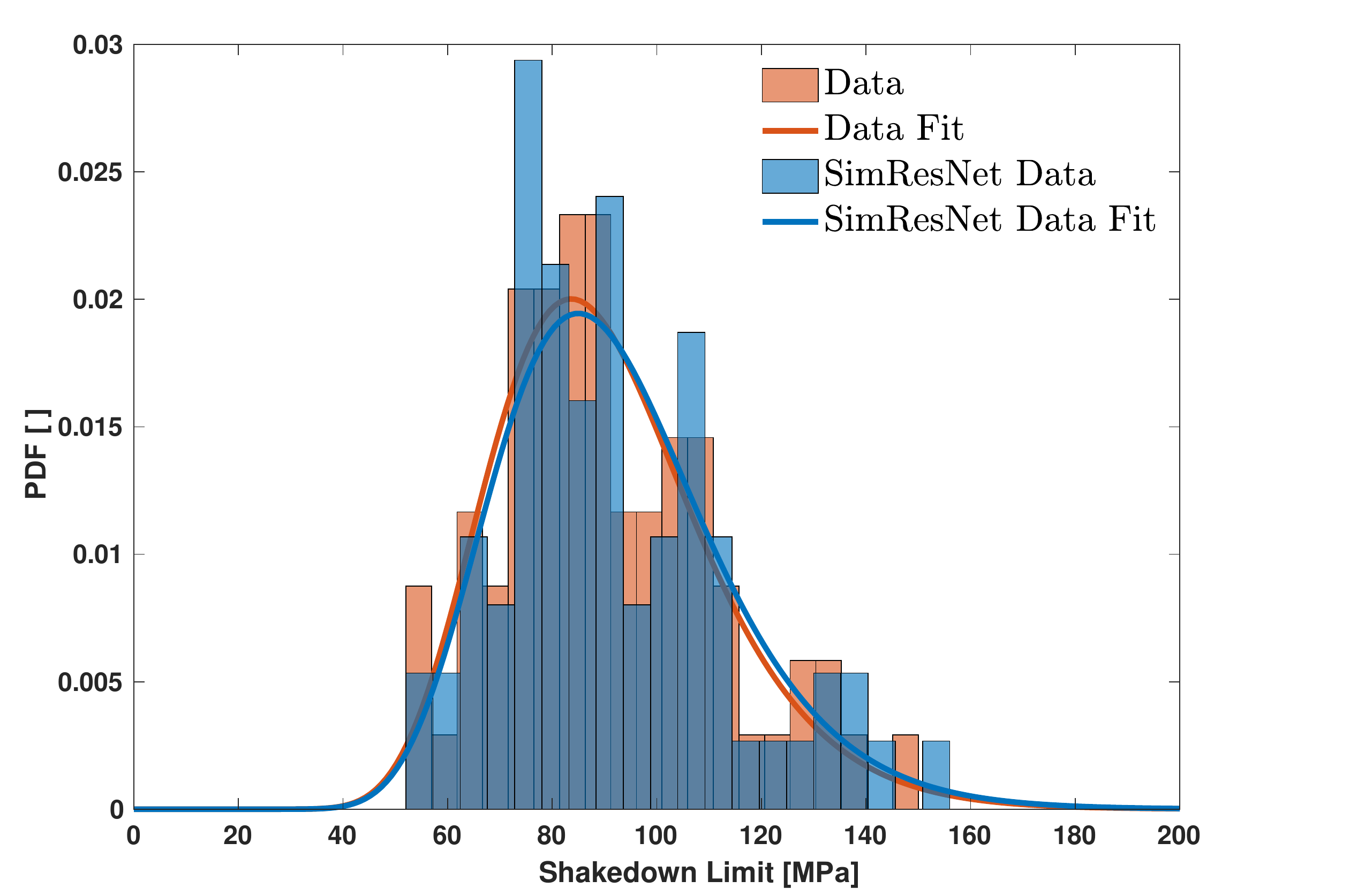}
\end{center}
\caption{Comparison of simulated data with the shakedown theorem and predicted data with the SimResNet. The data has been fitted by a lognormal distribution. The $L_1$ error of the fitted distributions is given by $0.06$.}\label{fig:comp_shk}
\end{figure}

Fig. \ref{fig:comp_shk} shows a comparison of simulated data with the shakedown theorem with predicted data of the SimResNet. The data points of the SimResNet were obtained by predicting a single shakedown limit for each micrograph of group V. For the training of the SimResNet, the maximum feret distribution of a single micrograph was utilized as input data. The fitted distribution of the SimResNet nearly matches the fit of the simulated data.

\clearpage
\section{Discussion and Conclusions}\label{conclusions}
The SimResNet allows fast predictions of microstructure-fatigue relationships, which is an important advantage for industrial applications. The simulation time of the shakedown limit of approximately 8 hours is compared to merely 40 seconds training of the neural network. No expertise in micromechanical modeling is required. The only prerequisite is that the shakedown theorem correctly depicts the fatigue behavior of ductile cast iron. However, the SimResNet can also be trained for use with other materials and models. Therefore we envision that the SimResNet could be used e.g. as a plugin in image analysis toolboxes.

The prediction of the shakedown limit by means of all graphite nodules of a micrograph is successful since the relative deviation between target and output is small. The error sum $\eta_j$ includes the individual error of each feature of the distribution to the target of the entire image. If one intuitively compares the error of the distribution, i.e. the mean value of all outputs with the target, the largest relative error of all determined SimResNets is only \SI{1.58}{\percent}. Since the error of the shakedown limit between output distribution and target are low for both data sets, it seems that a simple neural network like the SimResNet is flexible enough to handle complex relationships, such as microstructure-fatigue relationships. We have shown that the distribution of features influences the macroscopic shakedown limit. If the distribution of the feature is different from the distribution used in training, a different shakedown limit results. In contrast to group RN, the features of V are bimodally distributed. Thus, a micrograph of RN applied to V and vice versa leads to high errors, as shown in section \ref{results}.

It is notable that the predictions of the shakedown limit for all SimResNets of group V with V data are generally lower. The input distributions are more homogeneous in RN than in V. Because of the bimodal distribution of the features in group V, the SimResNet has more degrees of freedom. Moreover, for group V, a single graphite nodule could influence the shakedown limit more easily by violating the inequality constraint through high elastic stresses. On the contrary, in group RN, high hydrostatic stresses at spherical graphite nodules do not contribute to the von mises yield criterion used by the shakedown theorem. Thus, the shape parameters used for the SimResNet offer less flexibility.

By using combinations of two features, the prediction of the shakedown limit with SimResNet cannot be significantly improved. Depending on the features being linearly dependent, the SimResNet is not provided with additional flexibility. When the maximum feret increases the area of a graphite nodule increases as well. The aspect ratio of a graphite nodule is able to depict the morphology. However, the orientations of non-spherical graphite nodules to the load direction are statistically equally distributed but contribute to the input data of the SimResNet equally. Thus, this feature in fact might not provide much more in-depth data. The best results are obtained in the three feature case with improved training. This might be attributed to the high number of additional degrees of freedom of the SimResNet.

From the result that a single micrograph for the training is sufficient to make accurate predictions with the SimResNet, we conclude that the size of the micrograph is nearly representative for the microstructure of a group V or RN, respectively. When we use the SimResNet to predict a single shakedown limit for each micrograph of a group V or RN, the resulting distribution of shakedown limits fits very well with the distribution of the target (compare Fig. \ref{fig:comp_shk}).

To summarize, we applied the shakedown theorem to micromechanical models built from micrographs of cast iron. The simulation procedure was applied to micrographs containing non-spherical and spherical graphite nodules to determine the shakedown limit. The shakedown limit is a macroscopic stress level for which no failure due to accumulated plastic strain occurs. Moreover, we obtained morphological parameters through image analysis. A simplified SimResNet was trained with the morphological parameters as input and the shakedown limit as output target.

The main results are:
\begin{itemize}
	\item The SimResNet allows fast predictions of microstructure-fatigue relationships.
	\item The simplified architecture of the neural network (SimResNet) is capable of representing the complex relationships between microstructure and shakedown limit.
	\item The SimResNet provides better strength predictions for microstructures of cast iron with non-spherical graphite nodules. 
\end{itemize}

In the future we will work on the following open questions:

\begin{itemize}
	\item Apply the method to other microstructures and materials.
	\item Use of further data from experiments or alternative models for simulation of fatigue strength
\end{itemize}

\section*{Acknowledgments}
The work reported in this paper was partially funded by the Deutsche Forschungsgemeinschaft (DFG, German Research Foundation) under Germany's Excellence Strategy – EXC-2023 Internet of Production – 390621612. The material investigated originated from a project (IGF 18524N) funded through the German ministry for economy and technology (BMWi) by a resolution of the German Bundestag.
\clearpage
\bibliographystyle{elsarticle-harv}
\bibliography{SimNN.bbl}

\end{document}